\documentclass[prb,aps,epsf,twocolumn,showpacs]{revtex4}
\usepackage[dvips]{graphicx}
\usepackage{latexsym}
\usepackage{amsmath}
\begin{document}

\title{A synthesis of the phenomenology of the underdoped cuprates}

\author{T. Senthil and Patrick A. Lee}
\affiliation{ Department of Physics, Massachusetts Institute of
Technology, Cambridge, Massachusetts 02139}

\date{\today}
\begin{abstract}
The underdoped cuprates have a number of interesting and unusual properties that often seem hard to reconcile with one another. In this paper we show how many of these diverse phenomena can be synthesized into a single coherent theoretical picture. Specifically we present a description where a pseudogap and gapless Fermi arcs exist in the normal state above the superconducting transition temperature ($T_c$), but give way to the observed quantum oscillations and other phenomena at low temperature when the superconductivity is suppressed by a magnetic field. We show the consistency between these phenomena and observations of enhanced Nernst and diamagnetic signals above $T_c$. We also develop a description of the vortex core inside the superconducting state and discuss its relation with the high field phenomena.

\end{abstract}
\newcommand{\be}{\begin{equation}}
\newcommand{\ee}{\end{equation}}
\newcommand{\bea}{\begin{eqnarray}}
\newcommand{\eea}{\end{eqnarray}}
\newcommand{\bK}{\textbf{K}}

\maketitle

\section{Introduction}
In the last several years, a large number of experiments have studied the properties of
the underdoped cuprate materials. The phenomenology of these materials is rather unusual, and a unified theoretical picture of the diverse
phenomena is yet to emerge. First below a temperature scale $T^*$ there is the onset of the pseudogap in various probes. Remarkably Angle Resolved Photoemission Spectroscopy (ARPES) studies of the electron spectral function reveal the existence of gapless Fermi {\em arcs} in the pseudogap region above the superconducting transition\cite{hitcarpes}. The length of this Fermi arcs decreases with decreasing temperature (T), possibily extrapolating to zero\cite{kanigel} in the limit $T \rightarrow 0$. Second in a fairly wide range of temperatures above $T_c$, but below the $T^*$ line, there is an onset of local superconducting order without long range phase coherence\cite{corson,ong}. This is most strikingly evidenced by an enhancement of the Nernst effect and diamagnetic response in this region. A third piece of the phenomenology comes from recent quantum oscillation studies of the state obtained by suppressing
superconductivity in a high field at low temperature\cite{underdposc,suchitra}. Remarkably sharp oscillation frequencies are seen indicating the presence of a sharp Fermi surface with charge carriers that are fermionic Landau quasiparticles. The detailed behavior of the oscillation frequencies and other experiments have lead to a suggestion that the oscillations are associated with gapless electron pockets centered near the antinodal region of the Brillouin zone\cite{lebouef,fsrecnstrct,sudip,chenlee}.

Currently there is considerable confusion about how these various pieces of information about the normal ({\em i.e} non-superconducting) state of the underdoped cuprates fit together. The ARPES results at low field and high-$T$ seem to be in conflict with the existence of a closed Fermi surface at high fields and low-$T$. Further the suggested electron pocket interpretation of the oscillation experiments also has a difficulty with ARPES (and other probes at low field above $T_c$) - it is precisely along the antinodal region that the pseudogap is seen in the electron spectrum. It is puzzling how a relatively modest field of 40T can close the large anti-nodal gap. Finally the high field experiments (in particular studies of the Hall effect\cite{lebouef}) have questioned the correctness of the conclusions of Ong and co-workers based on their Nernst/magnetization experiments.

In this paper we show how these various seemingly conflicting results may be reconciled into a single coherent theoretical picture. We are cognizant of other interesting phenomena that have attracted attention recently, notably signatures of time reversal symmetry breaking\cite{bourges,greven,kapitulnik}. The role they play in determining the electronic structure is not understood, and we will not discuss them in this paper. Our theory is based on a small set of assumptions that we justify on various theoretical and empirical grounds. We then obtain a description of the underdoped cuprates where a pseudogap and gapless Fermi arcs occur above $T_c$ but give way to the observed quantum oscillation and other phenomena at high fields. Further we show that there is no real conflict between the conclusions of the Nernst/diamagnetism experiments and the high field oscillation experiments. As a bonus we obtain a description of the vortex core structure inside the superconducting phase, and discuss its relation with the phenomena described above.
In a companion paper\cite{tslee2} we explore a specific microscopic model that captures a number of aspects of the overall picture that emerges from the present one.

We begin by stating and justifying our central assumptions.
\begin{enumerate}
\item
{\em At zero field there is a coherence scale $T_{coh}$ below which the single particle electronic excitations become well defined in the sense of Landau, ie the inverse lifetime is less than $k_BT$. $T_{coh}$ increases with increasing hole concentration $x$ and is roughly of order $T_c$.}

The existence of such a coherence scale is a fairly general feature of metals near a Mott transition, and is supported by many theoretical calculations of a doped Mott insulator, for instance in slave boson theory\cite{kotliarliu} or through DMFT and their extensions\cite{dmft}. Empirically ARPES experiments see such a fairly sharp onset of coherent quasiparticles upon cooling into the superconducting state\cite{arpescoh}. Other indirect evidence comes from microwave transport\cite{optcoh}  and thermal Hall conductivity\cite{kappaxy} experiments where the scattering rate drops rapidly for $T$ below $\sim T_c$. 

\item
{\em The true ${H_{c2}}$ where local superconducting order is suppressed is large, possibly $\approx 150 T$ or higher for moderately underdoped materials. The resistive transition out of the superconductor however happens at a much lower field $H_{res}$. (For instance for $YBCO_{6.5}$,
$H_{res} \sim 40 T$).}

The large value of $H_{c2}$ is a central conclusion of the Ong Nernst/magnetization experiments. Thus we simply assume that this is correct, despite the recent criticism from Taillefer and co-workers. The $H-T$ phase diagram thus has the schematic form shown in Fig. \ref{hitcHT}. At low temperature and fields between the resistive transition and $H_{c2}$ the system may be dubbed a ``vortex liquid". A discussion of this assumption in terms of vortex structure will be given at the end of the paper.
We emphasize that the quantum oscillation experiments are done at very low $T$ at fields near $H_{res}$ and not near $H_{c2}$.
\begin{figure}
\includegraphics[width=8cm]{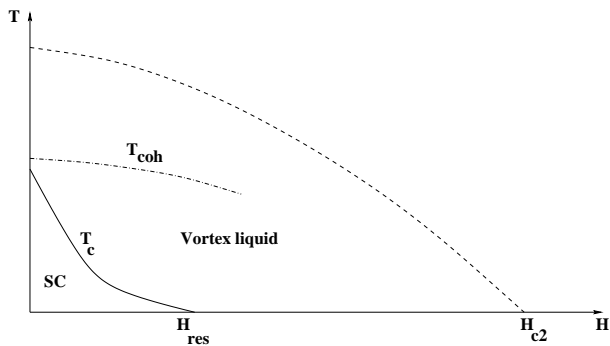}
\caption{Schematic field ($H$) - temperature ($T$) phase diagram showing the evolution of the superconducting properties. The full line is the resistive superconducting phase transition while the dashed line corresponds to the onset temperature of the crossover to local superconductivity. The coherence crossover temperature $T_{coh}$ is however not affected much by fields on the order of $H_{res}$. Not shown is the field induced magnetic ordering.}
 \label{hitcHT}
\end{figure}

\item
{\em Magnetic fields of order $H_{res}$ only destroy superconducting phase coherence but not the local superconducting pairing or the electron ``coherence".}

In other words $H$ of order $H_{res}$ is efficient in suppressing $T_c$ but not $T_{coh}$ or the pairing scale $\Delta$ (see Fig. \ref{hitcHT}). Not destroying the pairing gap is obvious from the previous assumption. Not destroying the electron coherence is an additional assumption. We note that in the simplest slave boson mean field description of the doped Mott insulator, $T_c$ and $T_{coh}$ are necessarily tied together. Thus the present assumption goes beyond the simplest slave boson mean field theory and must be justified theoretically in a more sophisticated theory.  Some empirical evidence in support of this assumption comes from STM studies of the vortex core in the superconducting state and will be discussed further at the end of the paper.

\item
{\em For magnetic fields $H > H_{res}$ but $\ll H_{c2}$, the Cooper pair phase has a finite memory time $\tau_{\phi}(H)$
which we assume is large compared to the inverse pairing gap $\Delta_0$, {\em i.e} $\frac{\hbar}{\tau_{\phi}(H)} \ll \Delta_0$. There is an associated memory length $\xi_{\phi} = v_F \tau_{\phi}$ where $v_F$ is the Fermi velocity.}

Thus $\xi_{\phi}$ is much larger than the `bare' coherence length $\frac{\hbar v_F}{\Delta_{0}}$ for $H \ll H_{c2}$. The scale $\tau_{\phi}$ is infinite for $H < H_{res}$ and decreases as $H$ increases toward $H_{c2}$. We expect that $\frac{\hbar}{\tau_{\phi}(H)} \sim \Delta_0$ for fields of the order $H_{c2}$. Further in the vortex liquid regime we expect that $\xi_{\phi}$ may be roughly estimated as the inter-vortex spacing, {\em i.e} $ \xi_{\phi} \sim \sqrt{\frac{hc}{eH}}$. In Fig \ref{hitcscgamma} we show a sketch of the expected form of the Cooper pair phase decay rate $\Gamma = \frac{1}{\tau_{\phi}}$ as a function of field $H$.

\begin{figure}
\includegraphics[width=8cm]{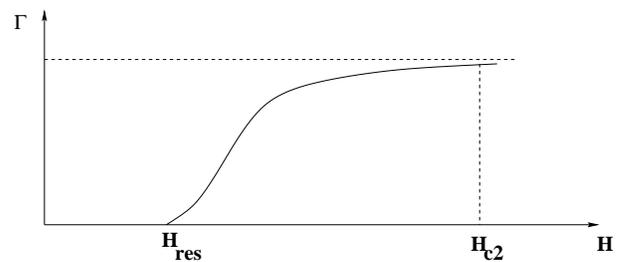}
\caption{Sketch of the Cooper pair phase decay rate as a function of field at $T = 0$. The saturated value at fields near $H_{c2}$ is of order $1/\Delta_0$.}
 \label{hitcscgamma}
\end{figure}

\item
{\em Finally we assume that fields of the order of $H_{res}$ induce a freezing of the dynamic incommensurate magnetic fluctuations known to exist in zero field.}

Field induced magnetic ordering has been previously reported in the LSCO family of cuprates\cite{lake}, and studied thoeretically\cite{subir}. Very recently direct evidence for field induced magnetic ordering has been obtained\cite{keimer} in $YBCO_{6.45}$. Further in $YBCO_{6.5}$
recent studies of quantum oscillations have shown that there is no Zeeman splitting of the oscillation frequencies\cite{qtmosczeeman},
as expected if there is magnetic ordering.  However the magnetic ordering appears as a low energy phenomenon that competes with the long range superconducting order. It can therefore be treated separately once the underlying electronic structure (including phenomena like the pseudogap) is established.

\end{enumerate}

With these assumptions we now examine the properties of the system in various regimes of temperature and field.

\section{Low $T$, fields $H \sim H_{res}$: emergence of large Fermi surface}
\label{emergeFS}

First consider low temperature $T \ll T_{coh}$ and fields in the range $H_{res} \sim H \ll H_{c2}$. To begin with we will also ignore the field induced magnetic ordering and incorporate it later. In this regime we may think in terms of an effective model of a superconductor disordered by quantum fluctuations of the order parameter phase. Previous studies of this model have suggested various possible ground states - the simplest to contemplate perhaps is a Cooper pair insulator. More exotic insulating states with fractionalized quasiparticles have also been proposed. Here we argue for the possibility of a particularly simple metallic state - namely just a Fermi liquid with a large Fermi surface. To appreciate this it is instructive to first consider the simpler problem of a clean $s$-wave BCS superconductor in $d = 2$ with short coherence length. Then the (Kosterlitz-Thouless) $T_c$ will be much smaller than the pairing gap $\Delta_0$, and can be readily suppressed in a magnetic field without suppressing the pairing. The resulting non-superconducting quantum ground state might of course be an insulator of Cooper pairs that breaks translational symmetry. However if it stays metallic the simplest option is to regain the large Fermi surface albeit with a reduced quasiparticle weight $Z$ at the Fermi surface. Below we demonstrate this directly for the $d$-wave case relevant to the cuprates through a simplified model.

Let $q^\dagger_{\alpha}$ create a low energy electron quasiparticle with spin $\alpha = \uparrow, \downarrow$ that exists below the coherence scale $T_{coh}$. At zero field in the SC state an appropriate low energy Hamiltonian is simply the BCS one:
\begin{eqnarray}
H_{eff} & = & \sum_{\textbf{K}} \epsilon_{\textbf{K}} q^\dagger_{\textbf{K} \alpha}q_{\textbf{K}\alpha} \nonumber \\
& + & \sum_{\textbf{K}} \Delta_{\textbf{K}} \left(q^{\dagger}_{\textbf{K}\uparrow}q^\dagger_{-{\textbf{K}\downarrow}} - q^\dagger_{{\bK} \downarrow}q^\dagger_{-{\bK} \uparrow}\right) + h.c
\end{eqnarray}
Here $\epsilon_{\bK}$ is the fully renormalized quasiparticle dispersion, and the pairing gap $\Delta_{\bK} \sim \Delta_0\left(Cos(K_x) - Cos(K_y)\right)$.
The operator $q^\dagger$ creates a `renormalized' low energy quasiparticle  that has a non-zero overlap $\sqrt{Z_0}$ with the bare electron. It is expected that $Z_0$ goes to zero as the hole concentration $x$ goes to zero. (For instance in the slave boson mean field theory, $Z_0 \sim x$.)

Now consider a situation where $\Delta$ maintains its phase only over a finite time scale $\tau_\phi$ and length $\xi_{\phi} = v_F \tau_\phi$. Then the pairing term in the Hamiltonian is modified to
\begin{equation}
H = \sum_{\textbf R} \hat{\Delta}({\textbf R}) \sum_{\textbf{r}} \eta({\textbf{r}}) \left(q^\dagger_{\textbf{R}\uparrow}q^\dagger_{\textbf{R+r}\downarrow}
- q^\dagger_{\textbf{R}\downarrow}q^\dagger_{\textbf{R+r}\uparrow} \right) + h.c
\end{equation}
where $\textbf{R}$ is a site of the lattice and the sum over $\textbf{r}$
extends over the four nearest neighbors. The constant $\eta(\pm
\textbf{x}) = + 1$, $\eta(\pm \textbf{y}) = - 1$. The amplitude
$\hat{\Delta}(\textbf{R})$ is written as \be \hat{\Delta}(\textbf{R}) =
\Delta_0 e^{i\hat{\phi}(\textbf{R})} \ee
where the phase $\hat{\phi}$ is to be regarded as a quantum operator-valued field.

At zero temperature, in space and imaginary time $\tau$ we take
the correlators of $\Delta(\textbf{R},\tau)$ to have the form \be
\langle \Delta^*(\textbf{R},\tau) \Delta(0,0) \rangle = \Delta_0^2
F(\textbf{R}, \tau)
\ee such that \bea F(\textbf{0}, 0) & = & 1 \\
F(|\textbf{R}| \rightarrow \infty , \tau) & \simeq &
e^{-\frac{|\textbf{R}|}{\xi_\phi}} \\
F(|\textbf{R}|, \tau \rightarrow  \infty) & \simeq & e^{-
\frac{|\tau|}{\tau_\phi}}. \eea
As a concrete and illustrative example we consider
the specific function \be
\label{modelFrs}
F(\textbf{R}, \tau) =
e^{-\frac{|\textbf{R}|}{\xi_\phi}}e^{- \frac{|\tau|}{\tau_\phi}}. \ee
This has the Fourier transform
\be
\label{modelF}
F(\textbf{p}, \omega) = \frac{16 \pi \xi_\phi^2}{(p^2\xi_\phi^2 + 1)^{\frac{3}{2}}}
\frac{\tau_\phi}{\tau_\phi^2 \omega^2 + 1}
\ee

 In the limit that the amplitude of the pairing gap $\Delta_o$ is
 small compared to the Fermi energy $E_F$ (which is not
 unreasonable for the cuprates), we may calculate the
 quasiparticle self-energy from scattering off the fluctuating
 pair order parameter in second order perturbation theory. This gives
 \be
 \Sigma(\textbf{K}, i\omega) = \Delta_{0K}^2 \int_{\textbf{p},\Omega} \frac{F(\textbf{p}, \Omega)}{i\left(\Omega - \omega \right) - \epsilon_{\textbf{p}- \textbf{K}}}
 \ee
 with $\Delta_{0K} = \frac{\Delta_0}{2} (Cos K_x - Cos K_y)$. For the specific functional form chosen for $F$ in Eqn. \ref{modelF} the integral can be evaluated readily both for small and large $|\omega|$ near the Fermi surface. For small $\omega$ and $\textbf{K}$ close to the Fermi surface we find
 \be
 \Sigma(\textbf{K}, i\omega) \simeq \frac{\Delta_{0K}^2}{\pi \Gamma^2}\left(-i\omega + v_F k_{\|}\right)
 \ee
 where $k_{\|}$ is the deviation of $\textbf{K}$ from the Fermi surface at the point of closest approach. For large $|\omega| \gg \Gamma$, we find
 \be
 \Sigma(\textbf{K}, i\omega)\simeq  - \frac{\Delta_{0K}^2}{i\omega + \epsilon_{\textbf{K}}}
 \ee
 For the electron Greens function this implies that for small $|\omega|$
 \be
 {\cal G}(\textbf{K}, i\omega) \simeq \frac{1}{\left(1 + \frac{\Delta_{0K}^2}{\pi \Gamma^2}\right)\left(i\omega - v_F k_{\|}\right)}
 \ee
 Thus the quasiparticle Greens function has a quasiparticle pole everywhere on the original large Fermi surface. However the quasiparticle weight has a reduction factor $Z_\Delta (\bK)$ due to the scattering off the pair field:
 \be
 Z_{\Delta} (\bK) =  \frac{1}{\left(1 + \frac{\Delta_{0K}^2}{\pi \Gamma^2}\right)}
 \ee
 We emphasize that this reduction is over and above any reduction $Z_0$ of the quasiparticle weight that is already present in defining the low energy effective model. The total quasiparticle weight therefore is $Z = Z_0 Z_\Delta$. Note that $Z_{\Delta}$ is strongly angle-dependent on the Fermi surface. Near the nodal direction $Z_\Delta \approx 1$ while near the antinodal direction $Z_{\Delta} \approx \frac{\pi \Gamma^2}{\Delta_{0}^2} \ll 1$. Thus the emergent large Fermi surface at low temperature has a highly anisotropic quasiparticle weight. Despite this strongly reduced $Z$ for the antinodal quasiparticle the scattering off the pair fluctuations does not cause a mass enhancement due to a compensation between the momentum and frequency dependent parts of the self energy of the $q$ field (at least
 within the model defined by the correlation function in Eqn. \ref{modelFrs}).

\begin{figure}
\includegraphics[width=8cm]{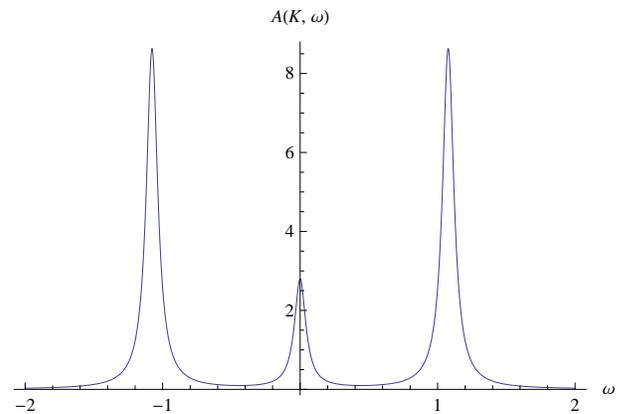}
\caption{Spectral function $A(\bK, \omega)$ at the antinodal point of the Fermi surface at low $T$ if the superconductivity is suppressed. We chose units where $\Delta_0 = 1$, and the parameter $\Gamma = 0.4$. A small broadening of the single particle energies $\gamma = 0.05$ (see Eqn. \ref{modelSgen}) was used in generating this plot.}
 \label{hitcspecfn_ANlowT}
\end{figure}

 For large $|\omega|$ ({\em i.e} $|\omega| \gg \Gamma$) on the other hand the form of the self energy implies that the
quasiparticle Greens function takes the form
\be
{\cal G}(\bK, i\omega) \simeq \frac{- \left(i\omega + v_F k_{\|}\right)}{\omega^2 + v_F^2 k_{\|}^2 + \Delta_{0K}^2}
\ee
which is the same as that inside the $d$-wave superconducting phase. Physically this just reflects the picture that the system looks like a superconductor at short time scales $< \tau_\phi$ and length scales $< \xi_\phi$. However at longer length and time scales the large Fermi surface emerges but with a reduced angle dependent spectral weight $Z$.
A simple expression for $\Sigma(\bK, i\omega)$ that interpolates between the small and large $|\omega|$ limits is
\be
\label{modelS}
\Sigma(\bK, i\omega) = \Delta_{0\bK}^2\frac{\left(-i\omega + \epsilon_{\bK}\right)}{\omega^2 + \epsilon_{\bK}^2 + \pi \Gamma^2}
\ee
Then the quasiparticle Greens function takes the form
\be
{\cal G}(\bK, i\omega) = \frac{1}{\left(i\omega - \epsilon_{\bK}\right)\left(1 + \frac{\Delta_{0\bK}^2}{\omega^2 + \epsilon_{\bK}^2 + \pi \Gamma^2}\right)}
\ee
This Greens function has a pole at zero frequency at the location of the Fermi surface and no zeroes anywhere in the Brillouin zone. The electron spectral function $A(\bK, \omega)$ may be readily extracted from this Greens function. In Fig \ref{hitcspecfn_ANlowT} we show a plot of $A(\bK, \omega)$ as a function of $\omega$ for $\bK$ at the antinodal portion of the Fermi surface.

Thus far we have ignored both  the field induced incommensurate magnetic ordering ({\em i.e} assumption 5 above), and the direct modification of the electron spectrum due to the magnetic field. The latter leads to the usual Landau quantization of orbits. The former can be incorporated in the low energy theory as a Hartree term
\be
H_m \sim gN_{\textbf{Q}} \sum_{\bK} \left(q^\dagger_{\bK + \textbf{Q}}\sigma^z q_\bK + h.c \right)
\ee
where $\vec N_{\textbf{Q}} = N_{\textbf{Q}}\hat{z}$ is the magnetic order parameter. Here the ordering wave vector $\textbf{Q} = 2\pi \left(1/2 \pm \delta, 1/2 \right)$ which is where soft fluctuations are seen in zero field. ($g$ is a coupling constant). This term will lead to a reconstruction of the Fermi surface into hole pockets centered near the nodal direction and electron pockets centered near the antinodal direction (see for instance Ref. \onlinecite{fsrecnstrct}).

A number of previous papers\cite{lebouef,suchitra,fsrecnstrct} have described how the resultant electron pockets can explain the frequency and other details of the quantum oscillation experiments, and we have nothing to add here. However our main point is that in the absence of the expected field induced magnetism the phase disordered $d$-wave superconductor has a large Fermi surface with an anisotropic quasiparticle weight as described above. The full Fermi surface is restored not by closing the pairing gap but by the appearance of quasiparticles with small weight in the anti-nodal region. We emphasize that the small quasiparticle weight does not directly affect the observability of the electron pocket in magnetization or resistivity oscillation experiments. {\em It is the emergence of this gapless large Fermi surface that makes the reconstruction into electron and hole pockets possible in the presence of magnetic order.} The crucial conceptual question then is to reconcile the emergence of this large gapless Fermi surface at low temperature with the pseudogap and other phenomena observed above $T_c$ at low fields. We turn now to this question.

\section{$T > T_c$, small $H$: pseudogap and Fermi arcs}

\begin{figure}
\includegraphics[width=8cm]{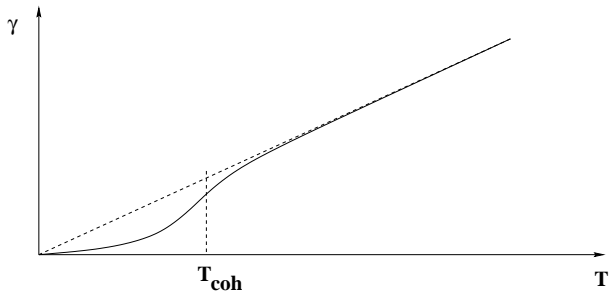}
\caption{Sketch of the single particle scattering rate $\gamma$ as a function of temperature.}
 \label{hitcspcoh}
\end{figure}

For $H = 0$, as $T$ increases above $T_c$, two things happen: (a) the superconducting phase coherence is lost and (b) the electron quasiparticle coherence is also lost. Thus the superconducting transition is not just a phase disordering transition but also a `coherence' transition for the electronic quasiparticle. We may model the loss of the superconducting phase coherence just as above with a fluctuating order parameter with phase coherence time $\tau_\phi$ and associated length scale $\xi_\phi$. We expect  $\xi_\phi$ to be roughly of order the spacing between thermally excited vortices. The Nernst/diamagnetism experiments imply that $\xi_\phi(T) \gg \xi_0 = \frac{\hbar v_F}{\Delta_0}$
at least for $T$ upto about $T_{onset} \approx 130 K$. Then the phase decay rate $\Gamma(T) \ll \Delta_0$ upto at least $T_{onset}$ though we may reasonably expect that it actually holds even upto $T^*$. The loss of single particle coherence may be modeled by introducing a scattering rate $\gamma$ which drops sharply for $T < T_{coh}$ and is large and roughly $\propto T$ for $T > T_{coh}$. This is depicted in Fig \ref{hitcspcoh}. The linear temperature dependence is not significant for the qualitative physics but affects details of the results below. In any case this is supported by experiments for $T$ above $T_{coh}$. Under these conditions the retarded Greens function for the $q$-field may be written
\be
G_R(\bK, \omega) = \frac{1}{\omega - \epsilon_\bK + i\gamma- \Sigma(\bK, \omega)}
\ee
(We emphasize once again that the measured electron Greens function will differ from the one above by a factor $Z_0   \sim o(x)$.)

The electron self-energy comes from scattering off the phase fluctuations of the superconducting order parameter. This may again be calculated as above. Taking the simple expression for $\Sigma$ used earlier in Eqn. \ref{modelS} to capture both low and high frequency regimes, we now find (in real frequency)
\be
\label{modelSgen}
\Sigma(\bK, \omega) = \frac{-\Delta_{0\bK}^2 \left(\omega - \epsilon_\bK + i \gamma\right)}{-\left(\omega + i\gamma \right)^2 + \epsilon_k^2 + \Gamma^2}
\ee
Let us examine this self energy in various regimes of interest. At low $T$ and zero fields $H = 0$, we have $\gamma = \Gamma = 0$. Then $\Sigma$ reduces to the usual $d$-wave BCS self energy. At low $T$, if the superconductivity is suppressed (in a field), we have $\gamma = 0$ but $\Gamma \neq 0$. Then the self energy reduces to the one described above in Eqn. \ref{modelS}, and yields a large Fermi surface. At high $T > T_{coh}$ and fields $H = 0$, we have $\gamma \gg \Gamma$. This is the regime where the pseudogap and the Fermi arc phenomena are seen in ARPES experiments. Writing Eqn. \ref{modelSgen} as
\be
\Sigma(\bK, \omega)  =  \frac{\Delta_{0\bK}^2}{\omega + \epsilon_\bK + i \gamma - \frac{\Gamma^2}{\omega - \epsilon_\bK + i \gamma}}
\ee
we see that the term involving $\Gamma$ in the denominator is small in this regime. Thus to leading order in the small number $\Gamma/E_0$ (where $E_0 = max(\gamma, |\omega|, |\epsilon_\bK|$), we simply have
\be
 \Sigma(\bK, \omega) \simeq  \frac{\Delta_{0\bK}^2}{\omega + \epsilon_\bK + i \gamma}
\ee
This is exactly the model self-energy introduced by Norman et al\cite{norman}, and Chubukov et al\cite{chubukov}, to describe ARPES data in the underdoped cuprates. In particular this captures well the essential features of the measured spectra. It leads to a pseudogap in the antinodal direction and a gapless Fermi arc near the nodal direction. The key point, as emphasized in Ref. \onlinecite{chubukov} is that so long as $\gamma > \sqrt{3}\Delta_{0\bK}$, the spectral function has a peak at zero frequency while if $\gamma < \sqrt{3}\Delta_{0\bK}$ it has particle-hole symmetric peaks away from zero energy. The former is the defining criterion for gaplessness in the experiments; near the nodal region the condition $\gamma > \sqrt{3}\Delta_{0\bK}$ is always satisfied above $T_{coh}$ in a finite segment of the Fermi surface. A gapless Fermi arc (as defined in the experiments) then follows. Furthermore the length of the Fermi arc scales linearly with the temperature.

\section{Relation to vortex structure}

It is useful to view the phenomenology described in the paper from the point of view of the structure of the vortex inside the superconducting state. It is interesting that STM experiments\cite{stmvrtx1,stmvrtx2,fischerrmp} find a large region (order $70 A$) around each vortex where the coherence peak is visible, but low energy `core' states appear inside the gap. The existence of the intermediate scale `halo'\cite{vortexhalo} where superconducting long range order is weakened but quasi-particle coherence remains is empirical support of our assumption 3.

\begin{figure}
\includegraphics[width=4cm]{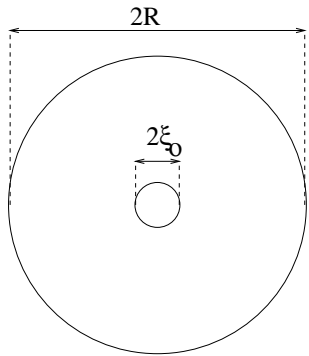}
\caption{Sketch of the vortex core showing the small inner core and the larger halo.}
 \label{hitcvrtxcore}
\end{figure}

From a theoretical point of view,
quite generally we know that there are two distinct energy scales $T_c$ and $\Delta_0$ in the underdoped cuprates.
We may then naturally expect two corresponding length scales $R = v_F/T_c$ and $\xi_0 = v_F/\Delta_0$
that may be visible in many properties including the structure of the vortex cores (see Fig. \ref{hitcvrtxcore}). Specifically there is a core of size $\xi_0$ and a `halo' of size $R \gg \xi_0$. Inside the halo the pairing suffers from strong phase fluctuations. This was first pointed out by Lee and Wen\cite{leewen} who suggested that the suppression of the superfluid density due to the excitations of nodal quasiparticles is responsible for the enhanced phase fluctuation in the halo region. However the presence of these two length scales for the vortex structure is fairly generic to superconductivity in a doped Mott insulator (see for instance Ref. \onlinecite{iofmil}).

The separation of scale between $H_{c2}$ and $H_{res}$ can  then be understood by recognizing that $H_{c2}$ and $H_{res}$ are the fields where the cores and the halos overlap respectively.
Clearly the quantum oscillation is probing the low temperature properties of overlapping halos.
In the case of $Bi-2212$ the tunneling spectrum exhibits peaks at $8 meV$ which shows roughly $4a \times 4a$ periodicity.  These peak structures are very naturally understood as being smoothly connected to the quasiparticle states of the full Fermi surface when the halos overlap above $H_{res}$. Specifically we may model the halo region as simply a disc of radius $R$ of the high field state embedded inside the bulk superconductor. The antinodal quasiparticle states of the high field state will then lead to peaks in the tunneling conductivity at a small gap that is induced by the non-zero superconducting order parameter that will exist in this halo region. This picture explains qualitatively an old empirical observation\cite{minigapscale} that the position of the low energy peak scales linearly with the maximum gap $\Delta_0$ (rather than with $\Delta_0^2$ as expected for core states in a BCS vortex).

Now consider $H<H_{res}$. As the temperature is raised, vortices and anti-vortices are created and $T_c$ is reached when their halos overlap. Above $T_c$ the overlapping halos form the vortex liquid smoothly connected to the $H>H_{res}$ region in Fig 1. It follows from our assumption 3 that the quasi-particles will exhibit coherence over a range of temperature $T_c < T < T_{coh}$.

\section{Discussion}

In conclusion, we see that a unified picture of the diverse phenomena in the underdoped cuprates can be constructed.
Here we briefly reiterate and discuss some key points of this paper. First we discussed quantum oscillation phenomena in high fields and low $T$ within the context of the assumption that local superconductivity is not killed in fields of order $H_{res}$. We showed that quasiparticle scattering off the pair field fluctuations resurrects a gapless large Fermi surface but with strongly anisotropic quasiparticle weight. Field induced incommensurate magnetic ordering can then reconstruct this large Fermi surface to give electron/hole pockets postulated in the literature. We reiterate again that the emergence for the gapless large Fermi surface is a precondition for the magnetic ordering to be able to produce the antinodal electron pockets.
Next we showed how the assumption of a coherence crossover scale $T_{coh}$ that is unaffected by fields $\sim H_{res}$ for the single particle excitations allows us to understand the pseudogap and Fermi arc physics above $T_c$ at low fields. Finally we discussed the structure of the vortex inside the superconducting state, and related it to various other phenomena.

In the future it should be extremely interesting for a variety of experiments to study the regime $T_c < T < T_{coh}$ at not too high fields. In this regime it may be possible to expose the emergence of the underlying large Fermi surface without magnetic ordering as described in Section \ref{emergeFS}. For instance STM studies may be possible in this region, and should see an enhanced zero bias conduction peak corresponding to the emergence of coherent quasiparticles even away from the Fermi arc region on cooling through $T_{coh}$. Transport experiments may also be a useful probe of this region though their interpretation will be complicated by the interplay between quasiparticle and vortex physics.
Another obviously important experiment is to study the vortex core structure through STM at low $T$ in underdoped samples, and follow the evolution of the halo and the core with decreasing doping.

It is interesting that we have been able to obtain a synthesis of the phenomenology without invoking broken symmetry
orders like staggered flux/d-density wave order\cite{lee,sudip}, or more exotic non-fermi liquid states (such as algebraic charge liquids\cite{kaul1,acl,galsubir}). Both of these possibilities have been considered in the literature. Indeed we have only invoked the incommensurate spin density wave induced by a field. There is good experimental support for such field-induced magnetic order. We emphasize however that even this field induced ordering is a low energy phenomenon. In particular
in our theory it plays very little role in electronic phenomena such as the pseudogap at high $T$.

A key challenge for microscopic theory is to describe the physics of the low coherence scale $T_{coh}$ for single particle excitations. In particular it is important to develop a microscopic theory for the incoherent scattering above  $T_{coh}$, the development of coherence at low temperatures below this scale, and the interplay with the pairing
into a spin singlet Cooper pair. In a companion paper\cite{tslee2} we present results from a particular model of a doped Mott insulator which captures these aspects of the physics.

\section*{Acknowledgments}
We are grateful to J.C. Davis, J. Hoffman, E. Hudson, N.P. Ong,  S. Sebastian, and L. Taillefer for many discussions that  greatly aided our thinking.  TS was supported by NSF Grant DMR-0705255, and PAL by NSF Grant DMR-0804040.

\end{document}